\documentclass{elsarticle}
\usepackage{lineno,hyperref}
\modulolinenumbers[5]

\biboptions{authoryear}

\begin{document}

\begin{frontmatter}

\title{High Resolution Imaging of the Magnetic Field in the central parsec of the Galaxy }

\author{  P. F. Roche$^{1}$ E. Lopez-Rodriguez $^{2}$,  C.M. Telesco$^{3}$, R. Sch{\"o}del $^{4}$, C. Packham$^{5}$  \\
$^{1}$ {Astrophysics, University of Oxford, DWB, Keble Road, Oxford OX1 3RH} \\
$^{2}$ {SOFIA Science Center,  NASA AmesResearch Center, CA 94035, USA} \\
$^{3}$ {Department of Astronomy, University of Florida, Gainesville FL 32611, USA} \\
$^{4}$ {Instituto de Astrofisica de Andalucia (CSIC), Glorieta de la Astronomia S/N, 18008 Granada, Spain}\\
$^{5}$ {Dept. of Physics \& Astronomy, University of Texas at San Antonio,1 UTSA Circle, San Antonio, Texas, 78249, USA } \\
}



\begin{abstract}
We discuss a high resolution (FWHM$\sim 0.45$ arcsec) image of the emissive polarization from warm dust in the minispiral in the Galactic Centre and discuss the implications for the magnetic field in the dusty filaments.  The image  was obtained at a wavelength of 12.5~$\mu$m with the CanariCam multimode mid-infrared imager on the Gran Telescopio Canarias.  It confirms the results obtained from previous observations but also reveals new details of the polarization structures.  In particular, we identify regions of coherent magnetic field emission at position angles of $\sim 45^o$ to the predominantly  north--south run of field lines in the Northern Arm which may be related to orbital motions inclined to the general flow of the Northern Arm.  The luminous stars that have been identified as bow-shock sources in the Northern Arm do not disrupt or dilute the field but are linked by a coherent field structure, implying that the winds from these objects may push and compress the field but do not overwhelm it.  The magnetic field in the  the low surface brightness regions in the East-West Bar to the south of SgrA* lies along the Bar, but the brighter regions generally have different polarization position angles, suggesting that they are distinct structures.   In the region of the Northern Arm sampled here, there is only a weak correlation between the intensity of the emission and the degree of polarization. This is consistent with saturated grain alignment where the degree of polarization depends on geometric effects, including the angle of inclination of the field to the line of sight and superposition of filaments with different field directions, rather than the alignment efficiency. 
\end{abstract}


\end{frontmatter}


\section{Introduction}

The central parsec of the Milky Way is a complex region hosting a range of unusual phenomena, including a supermassive black hole of mass $\sim 4 \times 10^6 $M$_\odot$, a young stellar cluster with numerous luminous mass-losing stars together with molecular and ionized gas filaments mixed with dust (e.g. \citealt{Genzel10}).  At mid-infrared wavelengths, the most prominent structure is the mini-spiral of gas and dust that comprises the Northern Arm and East-West Bar which lie within the 1.5--5 pc circumnuclear disk of molecular material around the Galactic Centre.  The mid-IR emission arises from dust grains heated to temperatures of $\sim$180--250~K (e.g. \citealt{Smith90, Gezari85}).  The derived dust temperature is fairly uniform in the bright filaments but there are local peaks close to the Northern Arm outflow sources.  Polarimetric measurements have revealed coherent polarization structures in the minispiral \citep{Aitken86, Aitken91, Aitken98, Glasse03}, which provide information on the magnetic field in the filaments.  The magnetic field appears to be quite ordered and strong in the Northern Arm, with estimated field strength of approximately 2~mGauss through indirect arguments, but much less ordered in the East-West Bar.  On the assumption that the dust grains have a net alignment to the field lines, determination of the polarization allows us to estimate the field directions in the plane of the sky.   \citet{Aitken98} have argued that the degree of polarization may be saturated in the Northern Arm, in which case, varations in the percentage polarization are related to the inclination of the field with respect to the plane on the sky.  Combination of polarimetric observations with radial velocities of the ionized gas then provides information on the 3-D structure of the filaments. 

Here we discuss  new higher resolution imaging polarimetry of the southern part of the Northern Arm and brightest region of the East-West Bar of the minispiral and the detailed correspondence between the polarization structures and other constituents in the central parsec.

 \begin{figure}
\vspace{-2cm}
\hspace*{-2.5cm}
\centering 
	\includegraphics[width=14cm]{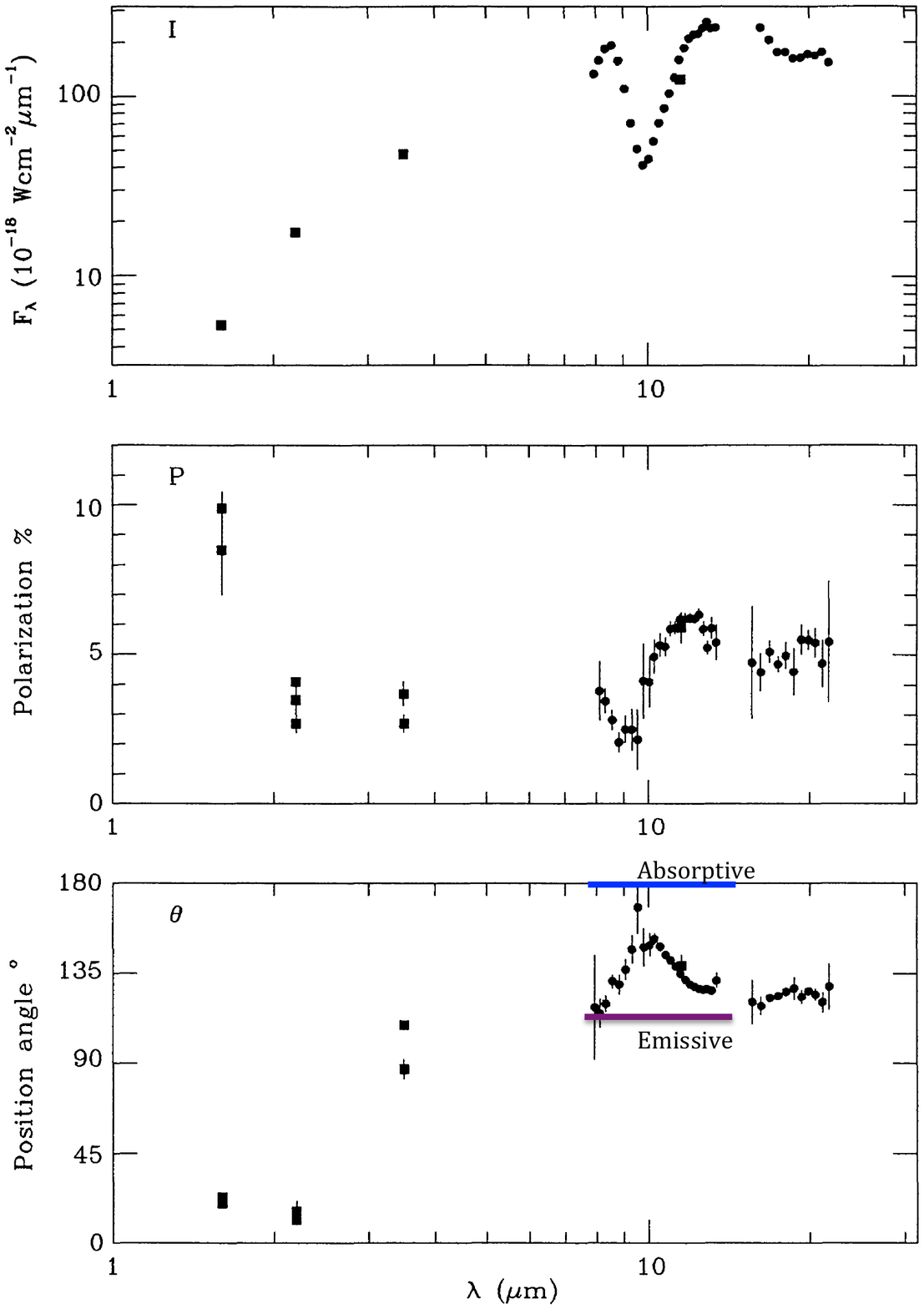}
	\vspace*{-3cm} 
	\caption{Measured infrared spectral dependence of polarization towards IRS1, the brightest compact source in the minispiral in the mid-IR.  From the top: Intensity, Percentage Polarization and Position Angle.  Note the change in position angle with increasing wavelength from the near- to the mid-IR reflecting the change from domination by interstellar  absorptive polarization to intrinsic emissive polarization.  Note also the rotation in position angle across the 10$\mu$m silicate band.  At short infrared wavelengths and near the peak of the silicate absorption feature, interstellar polarization dominates, but in the wings of the silicate band near 8 and 12 $\mu$m, the emissive component is dominant.  This behaviour is modelled by an emissive component at PA = 120 deg overlaid by an interstellar absorptive component at 0 deg.  The observations are made in apertures of $\sim4$~arcsec. Near-IR data are from \citet{Bailey84, Knacke77} while the 8-13 and 17-22~$\mu$m spectropolarimetric data are from \citealt{Smith00}. 
	}
	
	\label{fig:SpecPol}

\end{figure}

\section{Polarimetry of the Galactic Centre }

The centre of the Galaxy is viewed through an 8~kpc long column through the interstellar medium.  This renders it invisible at short wavelengths, but the effects of extinction are much reduced at infrared wavelengths.  It appears that the extinction over the central parsec is relatively uniform and characterised by a fairly constant extinction with an optical depth at the peak of the silicate absorption feature, $\tau_{9.7\mu{\rm m}} \sim3.6$ (e.g. \citealt{Roche85}).  Spectropolarimetry of the compact mid-IR sources in the minispiral has revealed that the polarization signatures can be interpreted by an intrinsic emissive polarization signature overlaid by a constant interstellar polarization component \citep{Aitken86}.   The polarization image presented here was measured at 12.5~$\mu$m where the emissive component is large while the interstellar absorptive component, which has a profile similar to the silicate absorption band, has decreased substantially from its peak near 10~$\mu$m.  The interstellar absorptive polarization is discussed in detail by \cite{Aitken91, Aitken98} who compare estimates of the interstellar polarization from mid-infrared spectro- and imaging- polarimetry and conclude that all of the results are consistent with a constant  absorptive polarization of ~1.8\% at PA= $5 \pm 5$ deg at 12.5~$\mu$m  arising from cool aligned grains in the ISM.  The observed spectral dependence of the infrared polarization measured towards IRS1 is shown in Fig. \ref{fig:SpecPol}, where the changes in position angle reveal the way the contributions from the emissive and absorptive components change with wavelength.  The detailed decomposition of the emissive and absorptive components between 8-13~$\mu$m can be found in \citet{Smith00}.

Because of dissipative forces in grains rotating in a magnetic field, non-spherical dust particles will tend to align so that the grain spin axis is parallel to the maximum moment of inertia.  The grain spin axis will precess around the magnetic field, providing a net alignment of an ensemble of dust grains, so that the long axis of a grain is preferentially aligned to be perpendicular to the magnetic  {\it B} field (see e.g. \citealt{Andersson15} for a review).  This results in polarized emission with the position angle normal to the  {\it B} field, so that when the emissive polarization vectors are rotated by 90$^{\rm o}$, they trace out the component of the field directions in the plane of the sky.

\section{Observations}
The observations were made with the Canaricam mid-IR imager mounted on the Nasmyth platform of the 10.4-m Gran Telescopio Canarias.  The data were obtained in the polarimetric mode wherein a rotating cryogenic half wave plate is used in conjunction with a focal plane mask and a Wollaston prism to produce simultaneous images of the e- and o-rays in three slots, each 2 x 20 arcsec \citep{Packham05}.  The telescope was stepped by 1 arcsec in Right Ascension after each exposure to provide full sampling and contiguous polarization maps of the minispiral.    The Galactic Centre only rises above 2 airmass for short periods above La Palma, and  the delivered image quality is seeing-limited at FWHM = 0.45 arcsec.  The polarization image was produced by coadding three sets of exposures which were aligned and corrected for the waveplate efficiency and for instrumental polarization induced by the off-axis reflection in the Nasmyth mirror   according to the prescriptions provided on the GTC/Canaricam website\footnote[1]{see http://www.gtc.iac.es/instruments/canaricam}.  The instrumental polarization induced by the tertiary mirror has been measured at 0.6\% with the position angle dependent on the instrument rotator angle. Observations of unpolarized stars taken with Canaricam (though not on the same nights as the  Galactic Centre) indicate residual instrumental polarization $<0.2\%$.  Finally, the image was corrected for interstellar polarization by taking  the measured polarization of the isolated luminous supergiant star, IRS3, which has a constant polarization position angle across the silicate band, to be representative of the interstellar component only (see the discussion in \citealt{Aitken98}).   The measured polarization of IRS3 in the current data is 1.7\% at PA $\sim$10 deg and we have used this value to correct for interstellar polarization. On the assumption that IRS3 is intrinsically unpolarised, this procedure corrects for any residual instrumental polarization as well as the interstellar absorptive component.  The plate scale of Canaricam is 0.08 arcsec pixel$^{-1}$. The data presented here have been smoothed to 0.24 arcsec pixel$^{-1}$ to improve the signal to noise ratio per point while preserving full sampling. For more details of the data analysis and processing see \citet{Roche18}.

\begin{figure*}
\vspace{-2cm}
\hspace*{-2.5cm}
\centering 
	\includegraphics[width=24cm]{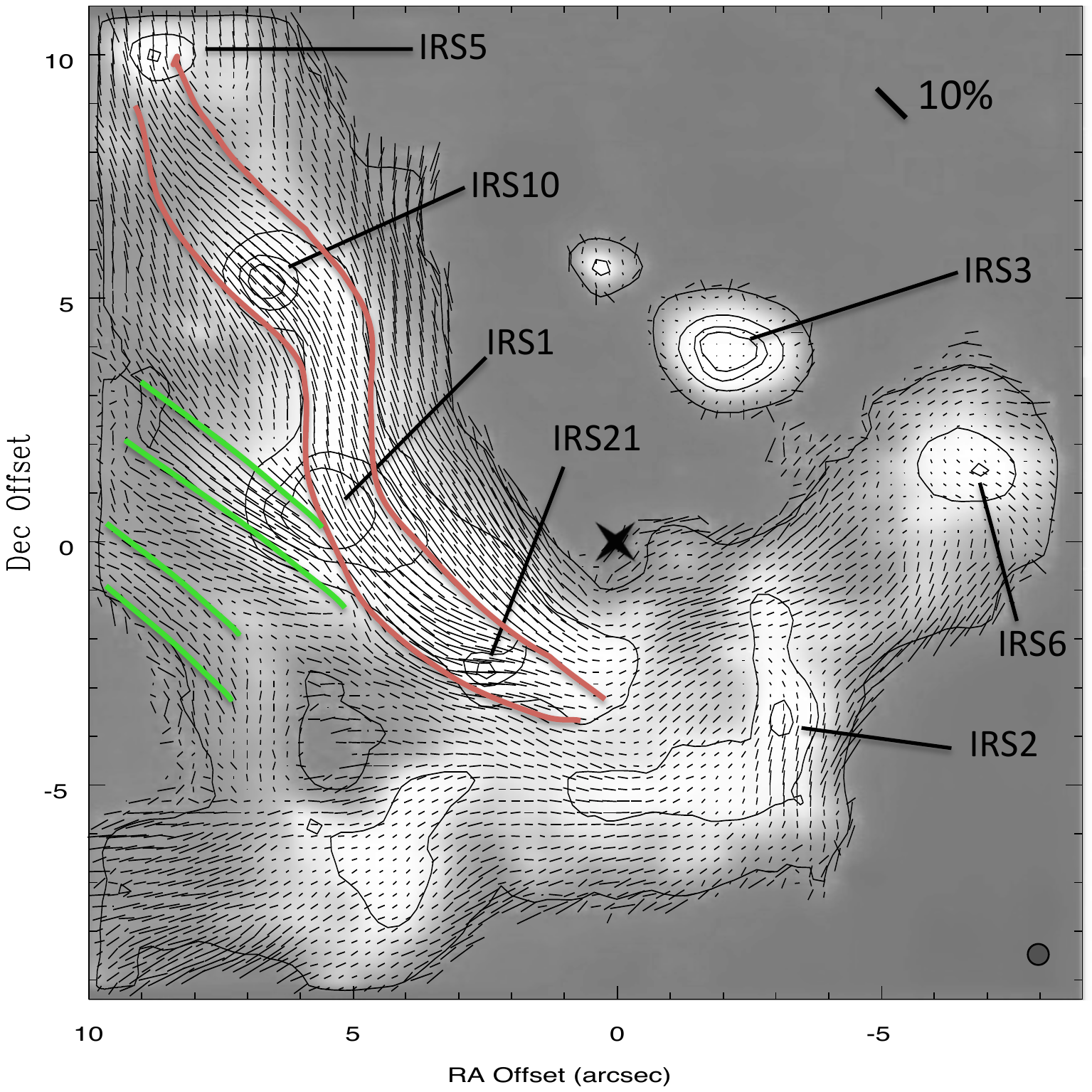}
	\vspace*{-1cm} 
	\caption{Polarization vectors superimposed upon a greyscale of the 12.5~$\mu$m intensity image. The vectors have been rotated by 90 deg so that they point along the magnetic field directions.    Coordinates are given as offsets in arcsec from the position of SgrA* which is marked by the cross.  The positions of some of the bright compact objects are indicated.  The lowest contour is at 1 Jy arcsec$^{-2}$ with further contours at $\sim$3, 6 and 9~Jy~arcsec$^{-2}$. The image size is indicated by the circle at the lower right.  Hand drawn lines indicate the run of coherent polarization vectors linking the embedded bow shock sources IRS5, IRS10, IRS1 and IRS21, while the lines to the south and east of IRS1 indicate the regions that cross the Northern Arm at  angles of $\sim45^o$.	In the East-West Bar, which runs from the south-east corner of the image to IRS6, the regions of lower intensity have polarization vectors pointing along the bar while the polarization in the brighter regions appear to arise from distinct field configurations.}
	
	\label{fig:Pol_map}

\end{figure*}

\section{Polarization image}

The emissive polarization map is shown in Fig. \ref{fig:Pol_map}.  In the image, the vectors have been rotated by 90 degrees, so that they point along the magnetic field directions.  It is clear that, as found previously, (e.g. \citealt{Glasse03}), the Northern Arm,   which is a coherent structure running from the north-east of the image before becoming indistinct south of SgrA* (e.g. \citealt{Irons12}) contains substantial regions of coherent vectors and that the bright compact sources neither dilute nor disturb the local field configurations.  This is seen more clearly and at higher resolution in the current image, where a run of vectors is seen linking the bright objects, IRS5, IRS10, IRS1 and IRS21.  These objects have been identified as luminous stars with substantial outflows which produce bow shocks where the stellar winds impact on the Northern Arm filaments  \citep{Tanner05, Sanchez14},   and produce local temperature peaks (\citealt{Smith90}).  The polarization at these positions closely follows that in the adjacent filaments indicating that the outflow sources are embedded in the magnetic fields and that while the outflow energy may compress and push the {\it B} fields, it does not disrupt them.  The coherent vectors linking the bright embedded outflow sources suggests that these objects have played a role in pushing the magnetic flux tube.  

In addition to this coherent structure linking the embedded objects and the general N--S vectors following the larger scale structure of the Northern Arm near the top of the image,  two regions of coherent vectors crossing the Arm at PA $\sim45^o$ can be identified to the east and south-east of IRS1. The first of these appears to converge or merge with the vector flow immediately south of IRS1 and continue to IRS21;  the second becomes indistinct where it meets a region of N--S vectors.  These {\it B} field structures may indicate regions where gas is flowing across the Northern Arm and in fact they provide a qualitative match to some of the Keplerian orbits proposed by  \cite{Paumard04}  and may indicate regions of inward gas flow.   

\begin{figure}
\vspace*{-2cm}
\hspace*{-1cm}
\centering 
	\includegraphics[width=14cm]{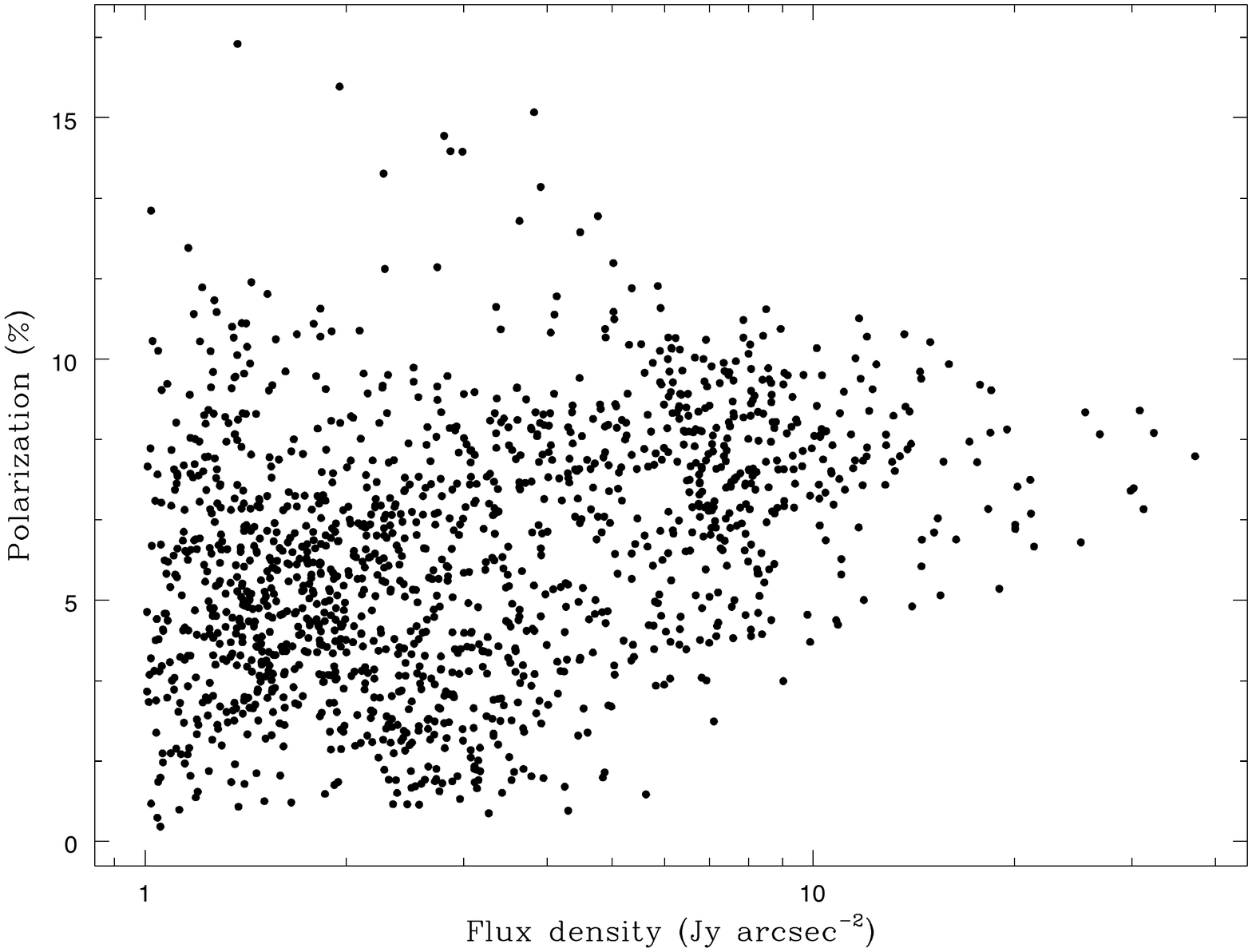}
	\vspace*{-1cm} 
	\caption{Polarization plotted against intensity in the region of the Northern Arm to the north and east of IRS21.  The intensity is plotted logarithmically to spread out the large number of points at the lower intensity levels.  
	}
	
	\label{fig:Pol_plot}

\end{figure}

Some grain alignment mechanisms, such as radiative torques acting on grains might be expected to lead to greater alignment and higher polarization in regions with high radiation fluxes (e.g. \citealt{Andersson15}). To further investigate this, the percentage polarization is plotted against the pixel intensity in the region of the Northern Arm to the north and east of IRS21 in Fig. \ref{fig:Pol_plot}.  Regions below the threshold for reliable polarization estimation (1 Jy arcsec$^{-2}$) are excluded.  The intensity varies by a factor of $>$30 in this region, peaking on IRS1, while the polarization ranges up to a maximum of  $\sim$14\%.   Such a plot has to be interpreted with caution as only the component of polarization in the plane of the sky is measured here. Furthermore, there are places where local minima appear to arise from the superposition of differently oriented filaments, and the distribution of the sources of luminosity  and the 3-dimensional structure of the filaments will all affect the measured polarization.   The best fit regression line is P(\%)= 4~S$^{0.23}$, where S is the flux density in Jy,  but the correlation coefficient is very low with R$^2$= 0.10.  Inspection of Fig.~\ref{fig:Pol_plot} suggests at most a modest increase in polarization with intensity suggesting that the effect expected from radiative torque alignment is modest  in the central region of the Milky Way.  However, \citet{Aitken98} proposed that the grain alignment in the Northern Arm may be saturated, in which case variations in the degree of polarization along the Northern Arm reflect changes in the orientation of the field with respect to the line of sight rather than the degree of alignment.  As several alignment mechanisms, including radiative torques (e.g. \citealt{Hoang16}) may be able to produce saturated alignment in grains with iron inclusions, this may not provide a robust test of the alignment mechanism.

At first sight, the {\it B} field structure in the East-West Bar seems to be much less ordered than that in the Northern Arm.  Inspection of the image shows that in the regions with lower emission intensity, the field lies predominantly along the Bar with PA $\sim~130^o$, but in contrast to the Northern Arm, the brighter regions in the Bar have polarization levels and position angles that diverge from the local diffuse emission.  This is especially apparent south of IRS2, where the polarization vectors are approximately north-south, and quite distinct from the underlying diffuse vector field which lies along the Bar. 	Similarly, near IRS6 at the western end of the Bar, there is a region where the polarization vectors have PA $\sim45^o$ close to the junction with the western arc. It seems that the bright regions in the Bar are not embedded in the general diffuse {\it B} field but are regions with distinct properties.  The region south of IRS2 shows strongly blue-shifted [Ne II] emission  \citep{Irons12} and has been considered as the southern extension of the Northern Arm where it re-emerges after curling below SgrA*;  the polarization vectors appear to be be consistent with that interpretation. Similarly, the region near IRS6 is where the western arc connects to or is superimposed on the Bar, and again we might expect distinct {\it B} field properties here. 

\section{Conclusions}

The 12.5~$\mu$m polarization image presented here reveals new details of the magnetic field in the central parsec of the Galactic centre.  The Northern Arm shows high levels of emissive polarization at 12.5~$\mu$m (peaking at 12\% in the region near and to the west of IRS1) with coherent field structures.   One of the most prominent magnetic structures links the four most luminous bow-shock sources in the Northern Arm, suggesting that the outflows play an important role in sculpting the field configuration in this region.  The embedded outflow stars probably compress and push the {\it B} field but do not overwhelm it.  Other coherent structures indicate regions where gas may flow across the Northern Arm and may represent regions of inflow.   In contrast, the bright regions in the East-West Bar have polarization vectors that are quite distinct from the underlying diffuse emission. It appears that the {\it B} field in the lower intensity diffuse regions lies along the Bar, but that the bright regions near IRS2 and IRS6 are distinct regions.
The intensity of the emission is only weakly correlated with the degree of polarization in the Northern Arm.  This is consistent with saturated grain alignment where the degree of polarization depends on geometric effects, including the angle of inclination of the field to the line of sight and superposition of filaments with different field directions, rather than the alignment efficiency.

\section*{References}
{}

\end{document}